\documentclass{article}
\usepackage{PRIME}
\usepackage[utf8]{inputenc} % allow utf-8 input
\usepackage[T1]{fontenc}    % use 8-bit T1 fonts
\usepackage{hyperref}       % hyperlinks
\usepackage{url}            % simple URL typesetting
\usepackage{booktabs}       % professional-quality tables
\usepackage{amsfonts}       % blackboard math symbols
\usepackage{nicefrac}       % compact symbols for 1/2, etc.
\usepackage{microtype}      % microtypography
\usepackage{lipsum}
\usepackage{amsmath}
\usepackage{amssymb}
\usepackage{enumitem}
\usepackage{mathtools}
\usepackage{fancyhdr}       % header
\usepackage{graphicx}       % graphics
\title{Predicting Tuberculosis from Real-World Cough Audio Recordings and Metadata
}

\author{
  George P. Kafentzis, Stephane Tetsing, Joe Brew, Lola Jover, Mindaugas Galvosas \\
  Hyfe Inc.\\
  U.S.A
  \AND
  Carlos Chaccour \\
  ISGlobal, Barcelona Institute for Global Health \\
  Spain
  \AND
  Peter M. Small\\
  University of Washington, Department of Global Health \\
  U.S.A
}

\begin{document}
\maketitle

\begin{abstract}
Tuberculosis (TB) is an infectious disease caused by the bacterium Mycobacterium tuberculosis and primarily affects the lungs, as well as other body parts. TB is spread through the air when an infected person coughs, sneezes, or talks. Medical doctors diagnose TB in patients via clinical examinations and specialized tests. However, coughing is a common symptom of respiratory diseases such as TB. Literature suggests that cough sounds coming from different respiratory diseases can be distinguished by both medical doctors and computer algorithms. Therefore, cough recordings associated with patients with and without TB seems to be a reasonable avenue of investigation. In this work, we utilize a very large dataset of TB and non-TB cough audio recordings obtained from the south-east of Africa, India, and the south-east of Asia using a fully automated phone-based application (Hyfe), without manual annotation. We fit statistical classifiers based on spectral and time domain features with and without clinical metadata. A stratified grouped cross-validation approach shows that an average Area Under Curve (AUC) of approximately $0.70 \pm 0.05$ both for a cough-level and a participant-level classification can be achieved using cough sounds alone. The addition of demographic and clinical factors increases performance, resulting in an average AUC of approximately $0.81 \pm 0.05$. Our results suggest mobile phone-based applications that integrate clinical symptoms and cough sound analysis could help community health workers and, most importantly, health service programs to improve TB case-finding efforts while reducing costs, which could substantially improve public health.
\end{abstract}

% keywords can be removed

\section{Introduction}
\label{sec:intro}

Tuberculosis (TB) is one of the leading causes of death worldwide, with an estimated 1.5 million deaths in 2020 alone~\cite{TBreport}. TB has a significant impact on economic development, has a disproportionate impact on vulnerable populations, and is increasingly due to drug resistant strains which are more difficult to treat. Thus, the eradication of TB is essential to save lives, reduce poverty, protect the most vulnerable, and safeguard against the spread of drug-resistant forms of the disease~\cite{Matteelli180035}.

A major impediment to TB eradication is the difficulty in finding cases. Approximately $40\%$ of people with TB are not diagnosed or reported to public health authorities because of challenges in accessing health facilities or failure to be tested or treated when they do. The development of low-cost, non-invasive digital screening tools would help to address this challenge. Cough has traditionally been used to identify people who may have TB and to monitor treatment~\cite{loudon1969cough, turner2015cough, fochsen2006health}. Advances in acoustic AI promises to passively detect and monitor cough and thus improve case finding~\cite{birring2008leicester, pham2016mobicough, liaqat2021coughwatch, zhu2022passive}. Furthermore, classification of coughs based on their sound may identify which individuals are most likely to have TB and thus help triage who should be prioritized for microbiological testing. Several previous studies have demonstrated the potential for cough sounds to be used to screen for TB~\cite{CODATB1, CODATB2, CODATB3, tracey2011cough, 9873469, 10007261, frost2022tb, pahar2022automatic}, though these were typically done in small samples and limited settings. Further development and evaluation of the diagnostic accuracy of AI algorithms for distinguishing tubercular from non-tubercular coughs are critical to move the field forward.

The CODA TB DREAM Challenge~\cite{CODA} is a major opportunity for advancing cough based diagnostics for TB. In brief, the Challenge collected data from people who presented to clinics across $7$ countries with new or worsening cough for at least $2$ weeks. Elicited coughs were recorded using the Hyfe Research app~\cite{hyfe}. Individuals were then comprehensively evaluated for TB with molecular and culture testing of their sputa. The Challenge made this data publicly available so that AI experts could use it to develop and test algorithms to predict TB status using features extracted from elicited coughs, either in the presence or absence of demographic and clinical factors used routinely for TB screening. The Challenge released a training data set that could be used to develop diagnostic algorithms. Groups who complied with the constraints of the Challenge submitted these algorithms to SAGE who evaluated their performance using a hold-out data set. In contrast, because the authors were involved in the data collection and excluded from participating in the official challenge, in this work we exclusively used the publicly available training data to develop and test our diagnostic algorithms.

Specifically, we attempted to predict TB using the training dataset made available in the CODA TB DREAM challenge, including (a) cough audio recordings (\textbf{Cough-only Experiment}) and (b) demographic and clinical metadata along with cough audio recordings (\textbf{Cough+Metadata Experiment}). Clinical data includes a list of variables that are allowed to be used in the prediction model, such as age, sex, heart rate, temperature, and others. Well-known machine learning algorithms were used for both tasks. Cough audio recordings were modeled using low-level descriptors (LLDs), i.e. features extracted from audio signals and represent basic properties of the sound, such as its pitch, loudness, and timbre. Some common examples of LLDs in audio analysis include spectral features that describe the energy distribution of a specific part of the audio waveform, such as spectral centroid, flatness, spread, and others. On the other hand, spectrotemporal features describe the distribution of energy over time across different frequency bands. Examples of features in this category include Mel-frequency cepstral coefficients (MFCCs), linear prediction coefficients (LPCs), log-mel spectrograms, and chroma features\cite{Muller2015}. Furthermore, temporal LLDs that describe time-domain characteristics of the audio signal, such as zero-crossing rate, energy, and amplitude envelope, are also used. Such feature sets have been successfully used in automatic cough detection~\cite{8857792, 8461394}. 

In Cough-only Experiment, we additionally used deep learning algorithms (two-dimensional convolutional neural networks - 2D-CNNs)~\cite{DLBook} operating on spectrotemporal features obtained from the cough audio signals. Different convolutional architectures are trained on the spectrotemporal representations in a similar manner to neural networks trained on images for tasks such as object detection~\cite{CNNobject} and image segmentation~\cite{CNNseg}. In audio processing, a spectrotemporal representation is suitable as an image-like input, as proved in many works~\cite{zhao2019speech, abdel2014convolutional, hershey2017cnn, vishnupriya2018automatic}. Such approaches have also been successfully applied in cough recognition~\cite{zhou2021cough, 8904554}. In Cough+Metadata Experiment, the presence of metadata in tabular form led us to an approach that trains models jointly with metadata and cough audio features. For this, we used conventional ML models (not CNNs) that are able to handle tabular data well. Areas under Receiver Operating Curve (ROC-AUCs) are provided as performance metrics, for two reasons: (a) ROCs are a convenient way to assess the trade off between sensitivity and specificity when projecting the use of a diagnostic tool in screening versus confirmation use cases, and (b) AUC evaluates the ability of a classifier to prioritize positive instances over negative instances in terms of ranking, compared to other measures such as accuracy, which assesses the correct identification of true and false positives based on a specific decision threshold. AUC is considered a more inclusive measure of performance, regardless of the threshold chosen.

The rest of the paper is organized as follows: Section 2 presents the details of the dataset. Section 3 discusses our approach for Cough-only Experiment while Section 4 does the same for Cough+Metadata Experiment. Finally, Section 5 discusses the results and Section 6 concludes this work.

\section{Dataset}
\label{sec:dataset}
The data are from health centers in $7$ countries (India, Philippines, South Africa, Uganda, Vietnam, Tanzania, Madagascar). The clinical research included the evaluation of all individuals who were 18 years of age or older and visited outpatient health centers for any health concern. Those who had a new or worsening cough that persisted for a minimum of two weeks were selected to participate in the study. 

During the initial visit, a survey was conducted to gather standard demographic and clinical information from the participants. Additionally, samples of sputum were collected for tuberculosis TB testing. As part of the study, the participants were requested to cough, and the cough sounds were recorded using the Hyfe Research app. The app guides the participants with a countdown (3-2-1) and prompts them to cough, capturing approximately half a second of the "explosive" sounds within a five-second timeframe. This process is repeated five times. Cough sounds identified as coughs by the Hyfe cough prediction algorithm are included for analysis. It is important to note that the number of solicited coughs may vary for each participant depending on how many times they coughed during each five-second recording interval. Moreover, the act of producing a solicited cough may trigger additional coughing, so the cough files in this dataset comprise a combination of solicited and spontaneous coughs. The sampling frequency of all recordings is $44100$ Hz but the signals are sub-sampled for each experiment (more information in the following sections). In total, $9772$ sounds were analyzed. A statistical review of the dataset is illustrated in Table~\ref{tab:stats}. 

\begin{table}[htb!]
	\caption{Statistical review of the dataset.}\label{tab:stats}
	\centering
	\resizebox{0.8\columnwidth}{!}{%
		\begin{tabular}{|l||c||c||c|}
			\hline
			\textbf{Metric}                                & \textbf{TB+} & \textbf{TB-} & \textbf{Total} \\ \hline
			Participants                                   & 297          & 810          & 1107           \\ \hline
			Total coughs                                   & 2930         & 6842         & 9772           \\ \hline
			Average number of coughs/participant {(}$\pm$ std{)} & 10.06 {(}$\pm$ 6.48{)}  & 8.65 {(}$\pm$ 5.15{)}  & 9.03 {(}$\pm$ 5.7{)}    \\ \hline
			Minimum number of coughs/participant           & 3           & 3             & -               \\ \hline
			Maximum number of coughs/patient               & 50           & 37             &  -              \\ \hline
			Total duration of coughs (minutes)             & 24.41        & 57.01        & 81.43          \\ \hline
		\end{tabular}%
	}
\end{table}

Moreover, in Table~\ref{tab:metadata} we present a list of demographic and clinical metadata used as features in Cough+Metadata experiment.

\begin{table}[htb!]
	\centering
	\caption{Demographic and clinical data used as features in Cough+Metadata experiment.}\label{tab:metadata}
	\resizebox{\columnwidth}{!}{%
		\begin{tabular}{|l||l||l|}
			\hline
			\textbf{Clinical or demographic datum} & \textbf{Description} & \textbf{Unit of measurement} \\ \hline
			Age                                    & \begin{tabular}[c]{@{}l@{}}Age calculated as date of collection - date of birth if known. \\ If date of birth is unknown, reported age at time of collection.\end{tabular} & Years                        \\ \hline
			Sex  & Sex at birth reported by participant   & Binary (male or female)   \\ \hline
			Height  & The height of the participant  & Centimeters    \\ \hline
			Weight   & The weight of the participant  & Kilograms    \\ \hline
			Reported duration of coughing          & {Self reported duration of current cough (days)}  & Days  \\ \hline
			Prior TB  & \begin{tabular}[c]{@{}l@{}}Self reported status of whether the participant had or\\ been told to have TB\end{tabular}    & Binary (yes or no)   \\ \hline
			Prior TB (Pulmonary)  & \begin{tabular}[c]{@{}l@{}}Self reported status of whether the participant had or\\ been told to have pulmonary TB\end{tabular}  & Binary (yes or no)    \\ \hline
			Prior TB (Extrapulmonary) & \begin{tabular}[c]{@{}l@{}}Self reported status of whether the participant had or\\ been told to have extrapulmonary TB\end{tabular}  & Binary (yes or no)  \\ \hline
			Prior TB (Unknown)   & \begin{tabular}[c]{@{}l@{}}Self reported status of whether the participant had or\\ been told to have neither pulmonary nor extrapulmonary TB\end{tabular}   & Binary (yes or no)    \\ \hline
			Hemoptysis  & \begin{tabular}[c]{@{}l@{}}Self reported status of whether the participant has ever\\ coughed blood\end{tabular}   & Binary (yes or no)   \\ \hline
			Heart Rate  & The heart rate of the participant measured at baseline  & Beats per minute    \\ \hline
			Temperature   & The temperature of the participant measured at baseline   & Celsius      \\ \hline
			Smoke last week   & \begin{tabular}[c]{@{}l@{}}Self reported status of whether the participant used combustible\\ tobacco and/or vaping products in the last 7 days.\end{tabular}   & Binary (yes or no)           \\ \hline
			Fever   & \begin{tabular}[c]{@{}l@{}}Self reported status of whether the participant had felt or experienced\\ fever in the last 30 days.\end{tabular}   & Binary (yes or no)    \\ \hline
			Night sweats  & \begin{tabular}[c]{@{}l@{}}Self reported status of whether the participant had experienced\\ fever in the last 30 days.\end{tabular}   & Binary (yes or no)    \\ \hline
			Weight loss   & \begin{tabular}[c]{@{}l@{}}Self reported status of whether the participant had experienced\\ weight loss in the past 30 days.\end{tabular}   & Binary (yes or no)   \\ \hline
		\end{tabular}%
	}
\end{table}

\section{Features}
\label{sec:feats}

\begin{figure}[htb!]
	\centering
	\includegraphics[scale=0.63]{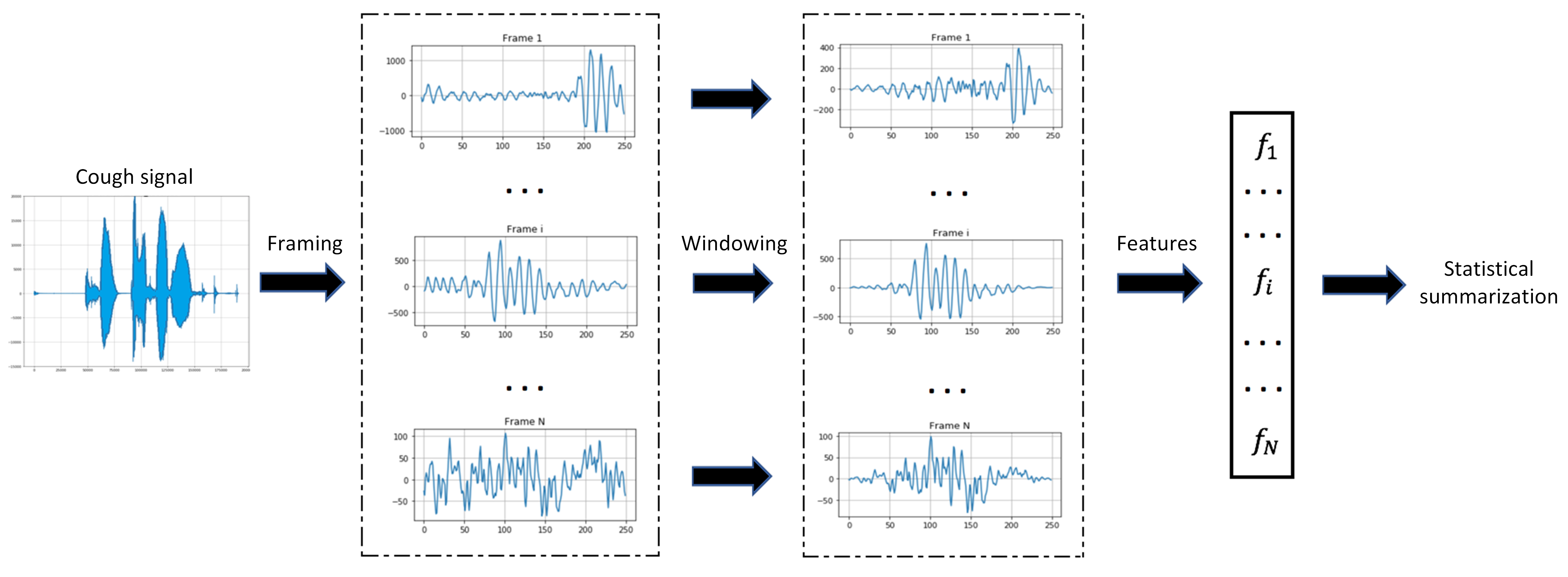}
	\caption{\it Feature Extraction Pipeline.}
	\label{fig:AFE}
\end{figure}

In this work we have utilized a series of audio features. The literature review suggests many different feature sets for compactly modeling sounds signals but most of these features were invented for audio, music, and speech signal use cases. Cough is a non-stationary signal, that is, it quickly changes over time. Thus we need to process it in short segments (the so-called frames) that are approximately stationary. Stationarity implies that temporal and spectral characteristics inside a frame do not significantly change. The length of each frame should be long enough to reliably estimate a feature set but short enough to ensure the measured features are representative of the whole frame. For these reasons, the usual frame length is around $20-50$ ms. For example, a $50$-ms window slides on the cough waveform with a step of $25$ ms (the so-called step or hop size or frame rate). This way, we have a $50\%$ overlap between successive frames. Finally, each frame is windowed (multiplied by a function in time that has desirable properties in the spectral domain) by a Hamming window. This process is illustrated in Fig.~\ref{fig:AFE}.
However, this approach leads to a large number of features per audio signal. Statistical summarization is a convenient way to reduce their number. 

\subsection{Low Level Descriptors}
To avoid the dimensionality curse~\cite{dimcurse, dimcurse2}, we would like to focus on features that are relatively low in number but sufficiently representative of the cough audio signal. There seems to be an agreement that temporal and spectral features (jointly termed as acoustic features) combined are useful up to a certain level~\cite{peeters2004large}. For our task we choose a set of LLDs consisting of temporal features like energy, Zero Crossing Rate, and Intensity, enriched by a more complex set of LLDs that includes Spectral measures (flux, entropy, 90\% rolloff, centroid, spread), and spectrotemporal measures such as Mel-Frequency Cepstral Coefficients (MFCCs) and log-filterbank spectrograms.

We will now briefly discuss the selected acoustic features and their properties that make them suitable for our task. In what follows, we consider an audio frame $x[n]$ and a chosen windowing function $w[n]$ with support in $[n-N, n+N]$.\\

\subsubsection{Temporal Features}

\begin{enumerate}[label=(\alph*)]
	\item Energy: the energy of a frame is a measure of the audio signals "size" or "strength", and is defined as
	\begin{eqnarray}
		E = \frac{1}{2N+1}\sum_{n=-\infty}^{+\infty}x[n]^2 w^2[n]
	\end{eqnarray}
	\item Zero-crossing Rate: the ZCR provides some rough information about the frequency distribution of the speech signal and is defined as the number of times a signal crosses the horizontal (time) axis: a high frequency signal should have a high number of zero crossings while a low frequency signal should have a low number of zero crossings. ZCR denotes the rate of sign-changes of the signal during the duration of a particular frame. The mathematical definition of ZCR is
	\begin{eqnarray}
		ZCR = \frac{1}{2N+1}\sum_{n=-\infty}^{+\infty}\frac{1}{2}\left|\mathrm{sgn}(x[n]) - \mathrm{sgn}(x[n-1])\right|x[n]w[n]
	\end{eqnarray}
	where $\mathrm{sgn}[\cdot]$ is the signum function.
	\item Intensity: the acoustic intensity of a sound is a physical quantity that can be defined as the average flow of energy (power) through a unit area measured in Watts per square meter. The human auditory system can detect a wide range of intensities, starting from $10^{-12}$ Watts per square meter and reaching up to $10$ Watts per square meter. These two extremes correspond to the threshold of hearing and the threshold of pain, respectively. Expressed in dB, the intensity of a signal $x[n]$ in air is defined as
	\begin{eqnarray}
		I = 10\log_{10}\frac{1}{P_0}\left[ \frac{1}{N}\sum_{n=-\infty}^{+\infty} (x[n]w[n])^2\right]
	\end{eqnarray}
	where $P_0$ equals $2 \times 10^{-5}$ Pa.
\end{enumerate}

\subsubsection{Spectral Features}

\begin{enumerate}[label=(\alph*)]
	\item Spectral Centroid: the spectral centroid (SC) is simply the center of gravity of the spectrum of a frame. It is computed considering the spectrum as a distribution which values are the frequencies and the probabilities to observe these frequencies are the normalized amplitude values. Let $X[k]$ be the N-point Fast Fourier Transform of the audio frame. Then the SC is defined as
	\begin{eqnarray}
		SC = \frac{\sum_{k=0}^{N-1}{kX[k]}}{\sum_{k=0}^{N-1}X[k]}
	\end{eqnarray}
	\item Spectral Spread: the spectral spread (SS) denotes the second central moment of the spectrum of the speech frame. The spectral spread describes the average deviation of the spectrum around its centroid, which is commonly associated with the bandwidth of the signal. Noise-like signals have usually a large spectral spread, while individual tonal sounds with isolated peaks will result in a low spectral spread. It is defined as
	\begin{eqnarray}
		SS = \sqrt{\frac{\sum_{k=0}^{N-1}(k-C)^2X[k]}{\sum_{k=0}^{N-1}X[k]}}
	\end{eqnarray}
	\item Spectral Roll-off: the 90\%-spectral roll-off is the frequency so that 90\% of the signal energy is contained below this frequency.
	\item Spectral Entropy: the Spectral entropy (SE) of a signal is a measure of its spectral power distribution. The concept is based on the Shannon entropy, or information entropy, in information theory. The SE treats the signal's normalized power distribution in the frequency domain as a probability distribution, and calculates the Shannon entropy of it. The Shannon entropy in this context is the spectral entropy of the signal. We can define the spectral entropy in terms of power spectrum and probability distribution of a signal. If $S[k]=|X[k]|^2$ is the power spectrum of an audio frame, then $P[k]$ is its probability distribution given by
	\begin{eqnarray}
		P[k] = \frac{S[k]}{\sum_{i=0}^{N-1}S[i]}
	\end{eqnarray}
	Finally, the SE is given by
	\begin{eqnarray}
		SE = - \frac{\sum_{m=1}^N P[m] \log_2 P[m]}{\log_2 N}
	\end{eqnarray}
	\item Spectral Flux: the spectral flux (SF) is a measure of how quickly the power spectrum of a signal is changing, calculated by comparing the power spectrum for one frame against the power spectrum from the previous frame. More precisely, it is usually calculated as the 2-norm (also known as the Euclidean distance) between the two normalized spectra. If $X_t[k]$ is the normalized power spectrum of frame $t$, then SF can be calculated as
	\begin{eqnarray}
		SF[k] = \left(\sum_{k=b_1}^{k=b_2} |X_t[k] - X_{t-1}[k]|^P\right)^{1/P}
	\end{eqnarray}
	where $b_1$ and $b_2$ are the band edges, in frequency bins, over which we calculate the SF. Usually, $b_1 = 0$ and $b_2 = N$, and $P=2$.
\end{enumerate}

\subsubsection{Spectrotemporal features}

\begin{figure}[htb!]
	\centering
	\includegraphics[scale=0.34]{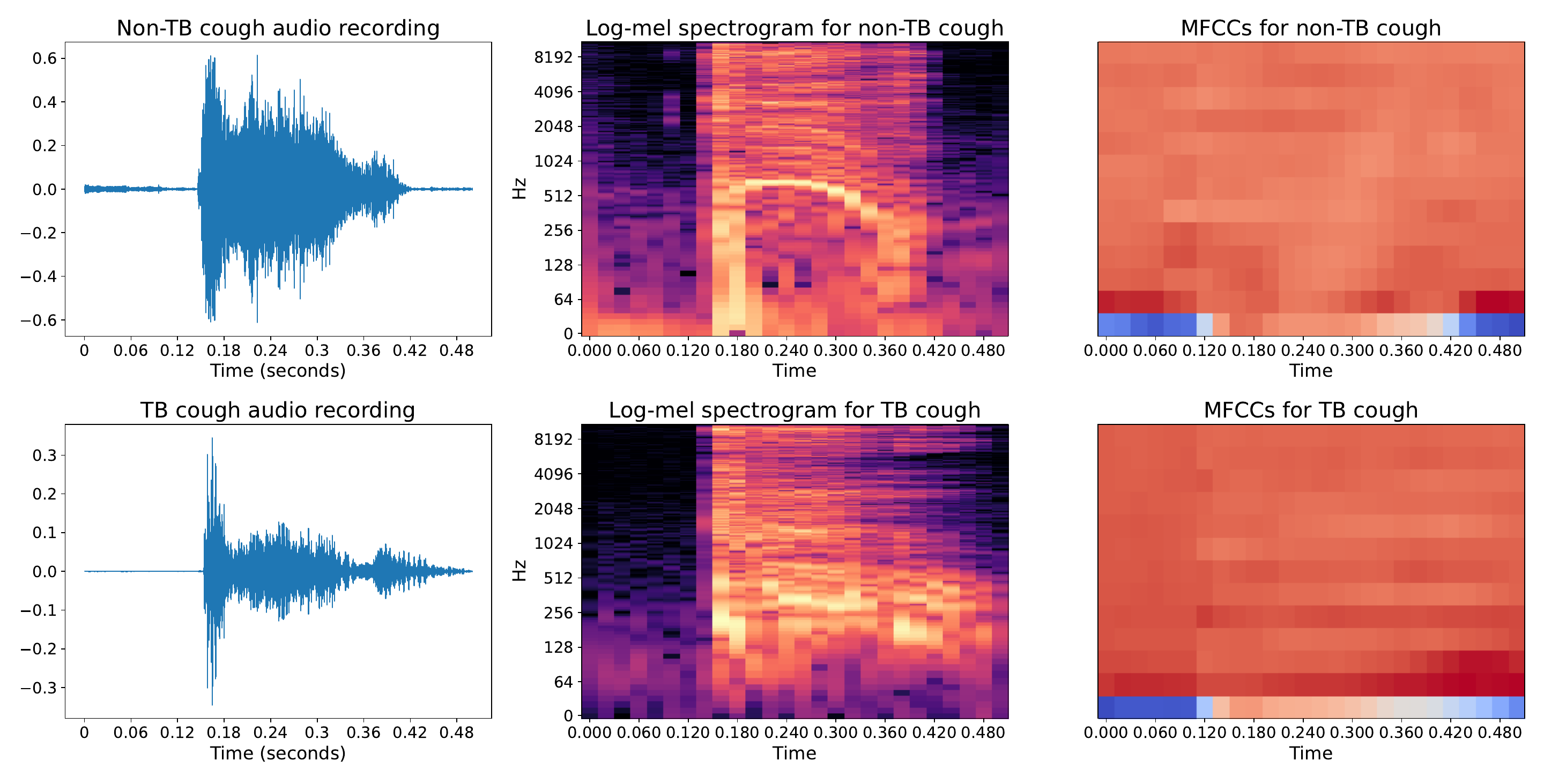}
	\caption{\it Audio recordings, log-mel spectrograms, and mel-frequency cepstral coefficients for two cough sounds, one from a TB patient (upper three panels), and one from a healthy patient (lower three panels).}\label{fig:logmel}
\end{figure}

\begin{enumerate}[label=(\alph*)]
	\item Log-mel spectrogram: The log-mel spectrogram is a specific type of spectrogram that combines two transformations: the Mel scale and the logarithmic compression. To create a log-mel spectrogram, the audio signal is first divided into short overlapping frames. Then, the Fourier transform is applied to each frame, resulting in a spectrogram. Next, the spectrogram is transformed to the mel scale using a filterbank that converts linear frequency values to mel scale values. This mel scale is defined as
	\begin{eqnarray}
		Mel(f) = 2595 \log\left(1 + \frac{f}{700}\right)
	\end{eqnarray}
	and is a perceptual scale of pitches judged by listeners to be equal in distance from one another. The log-mel spectrogram provides a concise representation of the audio signal that captures both frequency and temporal information.
	\item Mel-frequency Cepstral Coefficients: Similarly to the log-mel spectrogram, Mel Frequency Cepstral Coefficients (MFCCs) model the spectral energy distribution in a perceptually meaningful way, that is, MFCCs is a cepstral representation where the frequency bands are not linear but distributed according to the mel scale. Following computation of log-mel spectrogram, the Discrete Cosine Transform (DCT) is applied on the list of mel-log powers, as if it were a signal. The MFCCs are the amplitudes of the resulting signal.
\end{enumerate}
Parameters for generating all features are shown in Table~\ref{tab:mel} and an example of TB vs non-TB cough recordings, downsampled to $22050$ Hz, along with their corresponding spectrotemporal representations is depicted in Figure~\ref{fig:logmel}.

For standard ML models that require vectorized inputs, a set of statistics will be applied on these LLDs in order to summarize features from all frames of the audio signal. Let $v=\{x_1,x_2,x_3,\cdots,x_N\}$ denote a sample LLD consisting of $N$ values. We define:
\begin{enumerate}
	\item mean: the average value of an LLD
	\begin{eqnarray}
		\hat{x} = \frac{1}{N}\sum_{n=1}^N x_n
	\end{eqnarray}
	\item standard deviation: measures the amount of variation or dispersion of an LLD
	\begin{eqnarray}
		s = \sqrt{\frac{1}{N-1}\sum_{n=1}^N (x_n - \hat{x})}
	\end{eqnarray}
	%\item maximum: the maximum value of an LLD
	%\begin{eqnarray}
	%M = \text{max}\{x_n\}, \: \: n = 1, 2, \cdots, N
	%\end{eqnarray}
	%\item minimum: the minimum value of an LLD
	%\begin{eqnarray}
	%m = \text{min}\{x_n\}, \: \: n = 1, 2, \cdots, N
	%\end{eqnarray}
	\item skewness: measures the asymmetry of the sample distribution of an LLD about its mean
	\begin{eqnarray}
		b_1 = \frac{\frac{1}{N}\sum_{n=1}^N (x_n - \hat{x})^3}{\left[\frac{1}{N-1} \sum_{n=1}^N (x_n - \hat{x})^2 \right]^{3/2}}
	\end{eqnarray}
	\item kurtosis: measures the "tailedness" of the sample distribution of an LLD
	\begin{eqnarray}
		g_2 = \frac{(N+1)N}{(N-1)(N-2)(N-3)} \cdot \frac{\sum_{n=1}^N (x_n - \hat{x})^4}{\left[\sum_{n=1}^N (x_n - \hat{x})^2\right]^2} - 3\frac{(N-1)^2}{(N-2)(N-3)}
	\end{eqnarray}
\end{enumerate}
and these will be the final features for each audio signal. Reasons for such a selection are ease of implementation, either directly or by open-source software, robustness, and relevance to the task.

\begin{table}[htb!]
	\caption{\it Feature extraction parameters.}\label{tab:mel}
	\centering
	\resizebox{.7\columnwidth}{!}{%
		\begin{tabular}{|c|c|c|}\hline
			\multicolumn{3}{|c|}{\textbf{Parameters for temporal and spectral features}}  \\ \hline
			\multicolumn{1}{|c|}{\textit{\textbf{Frame Size}}} & \multicolumn{1}{|c|}{\textit{\textbf{Hop Size}}} & \multicolumn{1}{|c|}{\textit{\textbf{FFT size}}} \\ \hline
			\multicolumn{1}{|c|}{$0.05$ s}  & \multicolumn{1}{|c|}{$0.025$ s} & \multicolumn{1}{|c|}{$1024$}     \\ \hline
			\multicolumn{1}{|c|}{\textit{\textbf{Number of Features}}} & \multicolumn{1}{|c|}{\textit{\textbf{Window Type}}} & \textit{\textbf{Sampling Frequency}} \\ \hline
			\multicolumn{1}{|c|}{$62$ per frame}  & \multicolumn{1}{|c|}{Hanning}  & \multicolumn{1}{|c|}{$16000$} \\ \hline
			\hline
			\multicolumn{3}{|c|}{\textbf{Parameters for spectrotemporal features}}  \\ \hline
			\multicolumn{1}{|c|}{\textit{\textbf{Frame Size}}} & \multicolumn{1}{|c|}{\textit{\textbf{Hop Size}}} & \multicolumn{1}{|c|}{\textit{\textbf{FFT size}}} \\ \hline
			\multicolumn{1}{|c|}{$0.04$ s}  & \multicolumn{1}{|c|}{$0.02$ s} & \multicolumn{1}{|c|}{$2048$}     \\ \hline
			\multicolumn{1}{|c|}{\textit{\textbf{Number of Filters or Coefficients}}} & \multicolumn{1}{|c|}{\textit{\textbf{Window Type}}} & \textit{\textbf{Sampling Frequency}} \\ \hline
			\multicolumn{1}{|c|}{$128$ or $13$}  & \multicolumn{1}{|c|}{Hanning}  & \multicolumn{1}{|c|}{$22050$} \\ \hline
		\end{tabular}%
	}
\end{table}

\section{Experiments}
Implementations are carried out in Python~\cite{python}, (v. 3.9), using LibROSA~\cite{librosa} (v. 0.9.2), Tensorflow~\cite{abadi2016tensorflow} (v. 2.10), and Sci-kit Learn APIs~\cite{scikit-learn} (v. 1.22), while mathematical details can be found in~\cite{esl, DLBook}. A list of machine learning models and their hyperparameters (tuned or fixed) are shown in Table~\ref{tab:ML}. All classifiers are used for both experiments except for CNNs, which are only utilized in the Cough-Only experiment.

\begin{table}[htb!]
	\centering
	\caption{Machine learning models with their hyperparameters.}\label{tab:ML}
	\resizebox{\columnwidth}{!}{%
		\begin{tabular}{|c|l|}
			\hline
			\textbf{Algorithm}  & \multicolumn{1}{c|}{\textbf{Hyperparameters (constant and tuned)}}  \\ \hline
			Logistic Regression (LR)  & \begin{tabular}[c]{@{}l@{}}Solver: LBFGS\\ Penalty: L2\\ C: tuned from $10^{-5}$ to $1.0$\end{tabular} \\ \hline
			Support Vector Machine (SVM) & \begin{tabular}[c]{@{}l@{}}C: tuned from $0.01$ to $5$\\ $\gamma$: tuned from $10^{-1}$ to $10^{-5}$\end{tabular}  \\ \hline
			Multi-Layer Perceptron (MLP) & \begin{tabular}[c]{@{}l@{}}$\alpha$: tuned from $10^{-5}$ to $1.0$\\ Hidden layer size: $5$ \\ Neuron number: tuned from $256$ up to $4$ per layer, in decreasing order\\ Solver: Adam\\ Learning rate: $0.001$\\ Activation: ReLU\end{tabular}                                                         \\ \hline
			Random Forest (RF) & \begin{tabular}[c]{@{}l@{}}Estimators: tuned from $100$ to $500$\\ Maximum number of features: tuned between \textit{square root} and \textit{log2}\\ Maximum depth: tuned between $4, 6$, and $8$\\ Split criterion: tuned between \textit{Gini} and \textit{Entropy}\end{tabular}  \\ \hline
			AdaBoost (AB) & \begin{tabular}[c]{@{}l@{}}Number of estimators: tuned from $10$ to $500$\\ Learning rate: tuned from $10^{-4}$ to $1.0$\end{tabular}  \\ \hline
			\begin{tabular}[c]{@{}c@{}}CNN\\ (Cough-Only experiment)\end{tabular} & \begin{tabular}[c]{@{}l@{}}Convolution layers depth: tuned from $2, 3$ and $4$\\ Filter size: {[}$16, 32${]}, {[}$16, 32, 64${]}, {[}$16, 32, 64, 128${]} (first to last layer)\\ Kernel size = $3 \times 3$\\ Batch Normalization: tuned between yes and no \\
				(in between convolution and activation layers)\\ Pooling: tuned between Max and Average\\ Dropout: tuned between $0.25$ and $0.5$ (between dense layers)\\ Activation: ReLU\\ Number of dense Layers: $3$\\ Neuron number in dense layers: {[}$1024, 256, 128${]} \\
				(first to last layer)\end{tabular} \\ \hline
		\end{tabular}%
	}
\end{table}

\subsection{Cough-only Experiment}
\begin{figure}[htb!]
	\centering
	\includegraphics[scale=0.45]{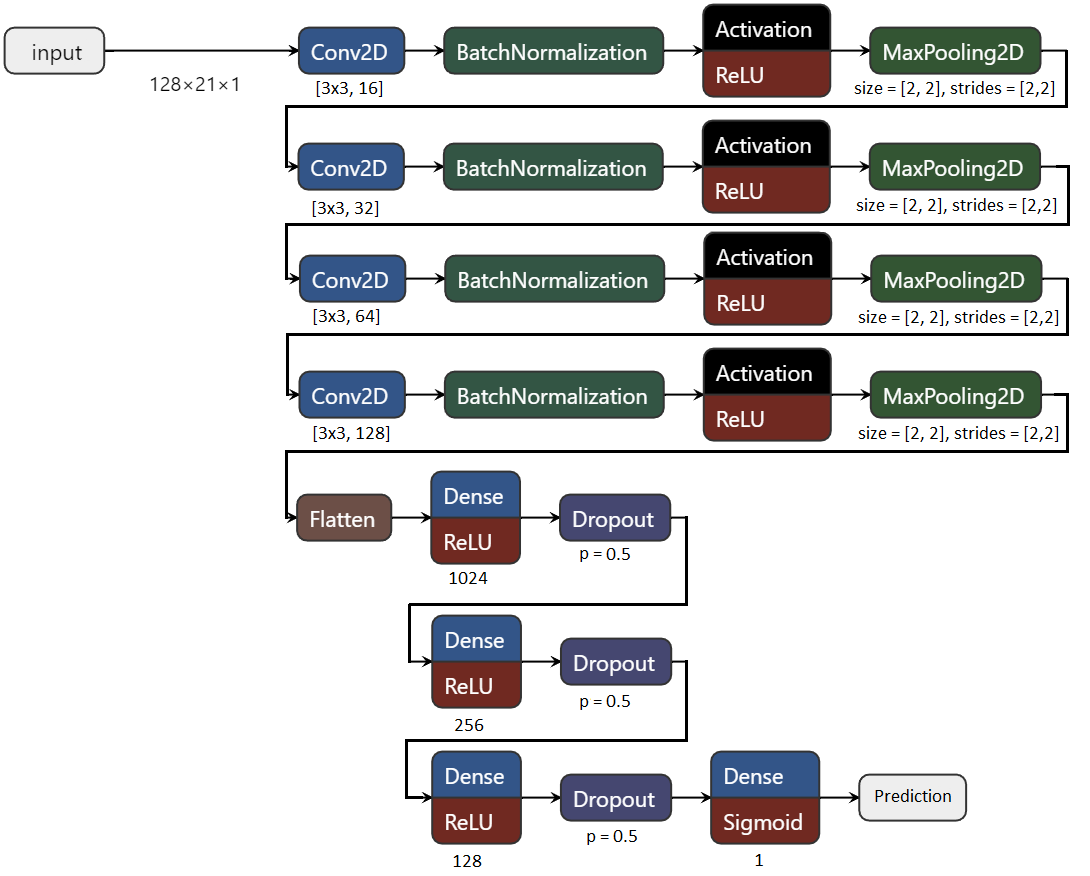}
	\caption{\it Best CNN architecture. $[MxN, L]$ under each Conv 2D block denote the filter size and the number of filters, respectively. $p=0.5$ under each Dropout block denotes the dropout probability while "size" and "strides" are parameters of the MaxPooling layer. Finally, numbers under Dense layers provide the number of neurons for each layer.}\label{fig:CNN}
\end{figure}

In the first experiment, TB prediction should be based on audio cough sounds only. Given the amount of available recordings ($9772$ files), deep learning approaches (CNNs) are also suitable for the classification task, along with conventional machine learning algorithms. The latter use high-dimensional vectors of temporal and spectral features as mentioned in Section~\ref{sec:feats}, while CNNs use spectrotemporal representations of cough sounds as input features. 
A $10$-fold stratified grouped cross-validation (CV) was selected to examine model performance. The proposed CV scheme ensures that coughs from each participant were either in the train set or the test set (and not in both). Different CNN-based architectures are tested but the one shown in Fig.~\ref{fig:CNN} is the one with the best average AUC results. Cough predictions are also averaged per participant.

Regarding training, a batch size of $32$ was chosen along with a number of $40$ epochs for training, a number small enough to prevent overfitting. Approaches with a validation set resulted in slightly worse performance, probably due to reducing the training set size. When training a model, it is often useful to lower the learning rate as the training progresses. We choose the Cosine Decay with restarts~\cite{loshchilov2016sgdr} scheduler that applies a cosine decay function with restarts to an optimizer step, given a provided initial learning rate (set to $0.0001$). It requires a step value (set to $4000$) to compute the decayed learning rate. The selected optimizer was Adam~\cite{kingma2014adam} with a binary cross-entropy loss function. 

\begin{figure}[htb!]
	\centering
	\includegraphics[scale=0.65]{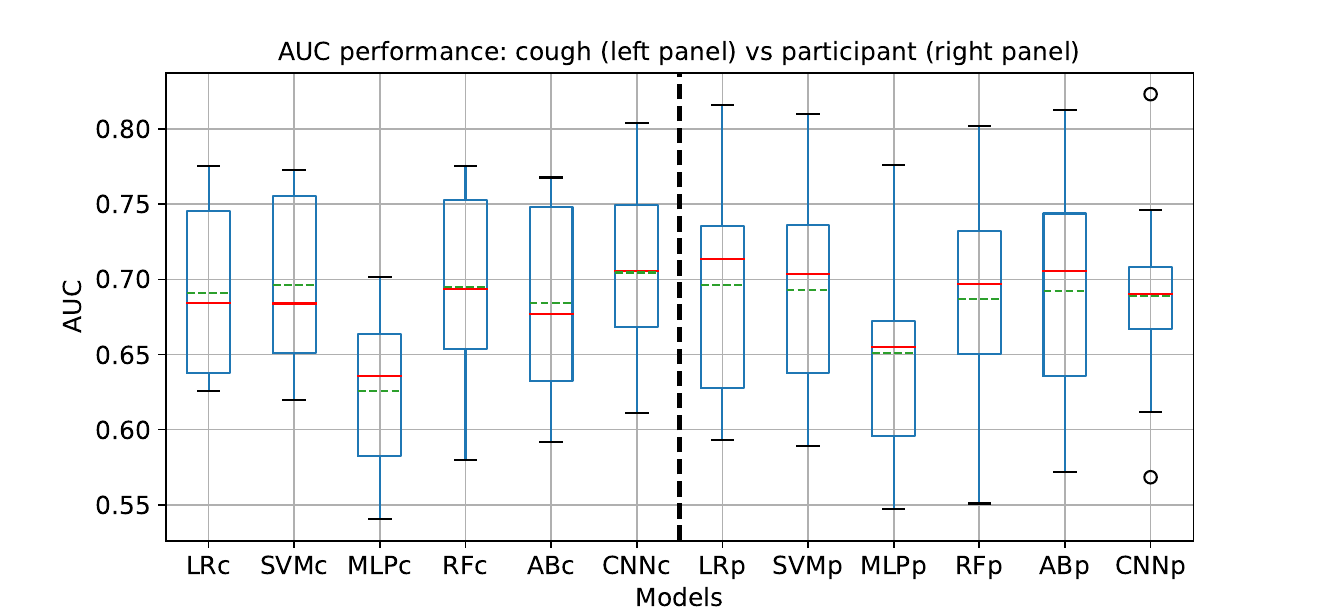}
	\caption{\it Cough-only experiment: boxplots of AUC values for all models. Left: per-cough assessment. Right: per-participant assessment. Red solid and green dashed lines denote mean and median value, respectively. For model abbreviations, see Table~\ref{tab:ML}.}\label{fig:CNNCV}
\end{figure}

In Figure~\ref{fig:CNNCV} we present a boxplot of AUCs averaged over a $10$-fold cross-validation, assessing both isolated cough sounds and aggregated per participant.
It should be noted that while the CNN outperforms other methods on average in isolated cough classification (AUC  $0.70 \pm 0.05$), when coughs are aggregated, a simple Logistic Regression achieves a higher average AUC ($0.69\pm 0.07$).

\subsection{Cough+Metadata Experiment}
In this experiment, we decided to jointly train a model on both the demographic and clinical metadata and the cough audio features. Since training 2-dimensional spectrotemporal data jointly with tabular data (metadata) is feasible but not straightforward, log-mel spectrograms and MFCCs were flattened to a 1-dimensional vector and concatenated to the rest of features in a tabular form.  

In this section, we briefly present the classification pipeline we follow in this experiment. Features are extracted, stacked, statistically summarized, and fed to the classifiers as inputs. Feature scaling is performed whenever necessary. We use stratified $5$-fold cross-validation to tune hyperparameters. Final predictions are made on several hold-out test sets, defined by a stratified, grouped $10$-fold cross-validation split. Hyperparameters were tuned in a nested stratified grouped $5$-fold cross-validation.  

\begin{figure}[htb!]
	\centering
	\includegraphics[scale=0.65]{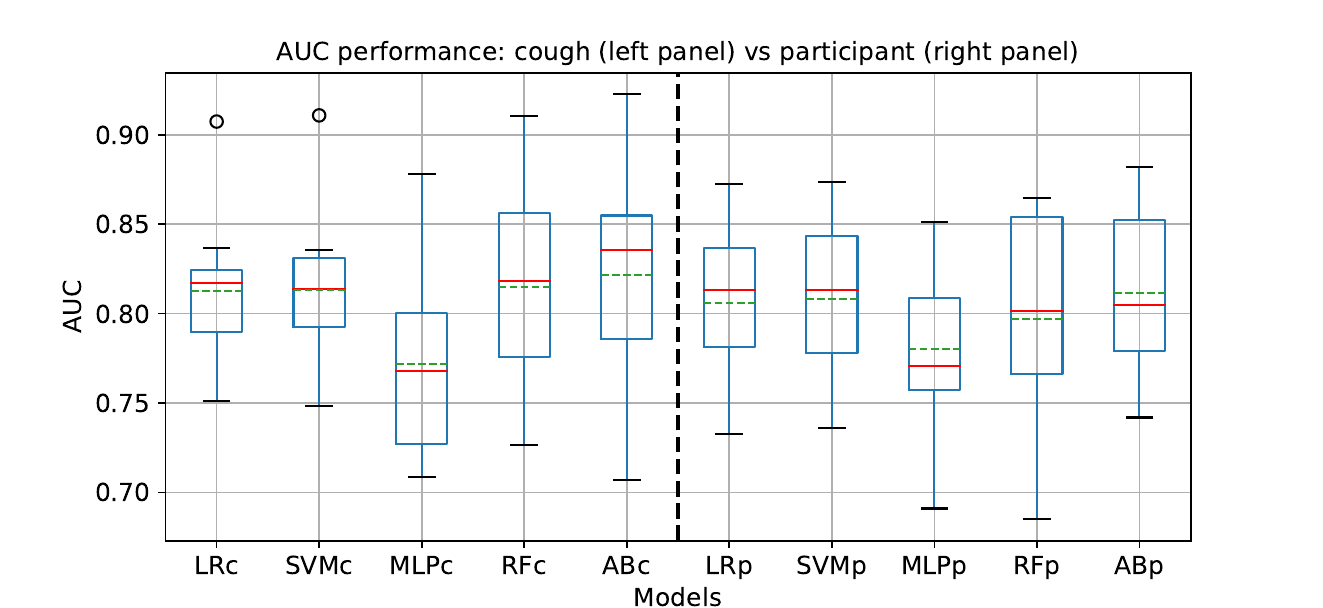}
	\caption{\it Cough+Metadata experiment: boxplots of AUC values for all models. Left: per-cough assessment. Right: per-participant assessment. Red solid and green dashed lines denote median and mean value, respectively. For model abbreviations, see Table~\ref{tab:ML}.}\label{fig:LRCV}
\end{figure}

In Figure~\ref{fig:LRCV} we present a boxplot of AUCs for our $10$-fold CV, assessing both isolated cough sounds and aggregated per participant.
An ensemble model, AdaBoost, achieves an average AUC of $0.82 \pm 0.05$ and an average AUC of $0.81 \pm 0.04$ is obtained per cough sound and per participant, respectively.

\section{Discussion}
Figures~\ref{fig:CNNCV} and~\ref{fig:LRCV} show that clinical and demographic data increase average AUC performance in both cough-based and participant-based classification. More specifically, in Cough-only experiment, none of the conventional ML methods showed a mean or median AUC greater than $0.70$ on a cough-basis classification. On the contrary, the CNN model achieved a mean and median AUC just above $0.70$. On a participant-basis classification, all models performed better on average, suggesting that aggregating probabilities might be beneficial to the prediction system. In Cough+Metadata experiment, AUC distributions are less spread out and both the average and median AUC is well above $0.80$ for all models but MLPs. It is interesting to note that Adaptive Boosting (AB) performs best on average in single cough classification but similarly to simpler models (LR, SVM) when aggregating cough probabilities. This behavior is consistent in other models as well, indicating that weighted averaging of probabilities might worth exploring. Overall, one may conclude that Multi-Layer Perceptrons (MLPs) seem to perform worst in all cases and Logistic Regression (LR), a simple, fast, and well-known statistical model performs quite well compared to more advanced models such as Adaptive Boosting (AdaBoost) and ensemble learning methods such as Random Forests (RFs).

\section{Conclusions and Future Work}
\label{sec:conc}
In this work, we present Hyfe's approach for diagnosing TB based solely on the publicly available training data from the CODA DREAM Challenge. We demonstrate the potential of both deep learning approaches and conventional features from audio processing as predictors for tuberculosis, along with standard demographic and clinical metadata. Results are encouraging, especially for models trained on both audio and tabular data. To our knowledge, this is the first study that compares machine learning techniques on such a diverse, large, automatically collected dataset of cough audio recordings. 
Importantly, our results on Cough+Metadata experiment are particularly relevant to TB control programs given the complexity and poor performance of conventional sputum microscopy~\cite{davies2008diagnosis}. If confirmed in larger community-based studies, our results suggest that the accuracy of cough sounds is sufficient for triaging which coughing patients should be prioritized for definitive TB diagnostic evaluation. 
This effort was not exhaustive and there are many improvements and alternatives which might further improve the audio analysis, starting from enriching the feature set and up to selecting different classifiers and feature selection schemes that have not been discussed in this work. In addition, there remain questions such as the optimal number of features, the most appropriate features that can model cough sounds adequately, the best classifier in terms of complexity, convergence speed, and accuracy, and the parameters of the analysis, among others. Moreover, the dataset used in his study provides a baseline dataset for researchers and engineers interested in developing statistical learning methods for TB prediction.
These preliminary results suggest that mobile phone-based apps that integrate clinical symptoms and cough sound analysis could help community health workers to identify TB patients. If integrated into digital health systems~\cite{nsengiyumva2018evaluating, mitchell2019digital} of low-resource countries with a high burden of TB, such as Nikshay~\cite{nikshay, kumar2020nikshay, dey2020awareness} in India, such devices would also allow TB control programs to better understand the epidemiology and clinical management of those suspected of having TB. 

\paragraph{Acknowledgments}
The datasets used for the analyses described were contributed by Dr. Adithya Cattamanchi at USCF and Dr. Simon Grandjean Lapierre at University of Montreal and were generated in collaboration with researchers at Stellenbosch University (PI Grant Theron), Walimu (PIs William Worodria and Alfred Andama); De La Salle Medical and Health Sciences Institute (PI Charles Yu), Vietnam National Tuberculosis Program (PI Nguyen Viet Nhung), Christian Medical College (PI DJ Christopher), Centre Infectiologie Charles Merieux Madagascar (PIs Mihaja Raberahona \& Rivonirina Rakotoarivelo), and Ifakara Health Institute (PIs Issa Lyimo \& Omar Lweno) with funding from the U.S. National Institutes of Health (U01 AI152087), The Patrick J. McGovern Foundation and Global Health Labs.

%\printbibliography %Prints bibliography
%\bibliographystyle{PRIME}
\bibliographystyle{IEEEtran}
\bibliography{refs}

\end{document}